\documentclass[useAMS,usenatbib]{mn2e}

\usepackage{graphicx}
\usepackage{aas_macros}

\title[TTS Magnetic Loops]{Mechanical Equilibrium of Hot, Large-Scale Magnetic Loops on T Tauri Stars}

\author[A. N. Aarnio, J. Llama, M. Jardine, and S. G. Gregory]
{A. Aarnio$^{1}$\thanks{E-mail: aarnio@umich.edu}, J. Llama$^{2}$, M. Jardine$^{2}$, and S. G. Gregory$^{3}$\\
$^{1}$University of Michigan Astronomy Department, 500 Church Street, Ann Arbor, MI 48109 USA\\
$^{2}$SUPA, School of Physics and Astronomy, University of St Andrews, St Andrews, KY16 9SS, UK\\
$^{3}$California Institute of Technology, MC 249-17, Pasadena, CA 91125 USA}
\begin{document}

\date{Submitted Nov 2011}

\pagerange{\pageref{firstpage}--\pageref{lastpage}} \pubyear{2011}

\maketitle

\label{firstpage}

\begin{abstract}
The most extended, closed magnetic loops inferred on T Tauri stars confine hot, X-ray emitting plasma at distances 
from the stellar surface beyond the the X-ray bright corona and closed large-scale field, distances comparable to 
the corotation radius. Mechanical equilibrium models have shown that dense condensations, or ``slingshot prominences'', 
can rise to great 
heights due to their density and temperatures cooler than their environs. On T Tauri stars, however, we detect 
plasma at temperatures hotter than the ambient coronal temperature. By previous model results, these loops should 
not reach the inferred heights of tens of stellar radii where they likely no longer have the support of the external 
field against magnetic tension. In this work, we consider the effects of a stellar wind and show that indeed, hot 
loops that are negatively buoyant can attain a mechanical equilibrium at heights above the typical extent of the 
closed corona and the corotation radius.
\end{abstract}


\section{Introduction}

\begin{figure}
\begin{center}
\includegraphics[width=85mm,clip=true,trim=30.0 0 10. 20.]{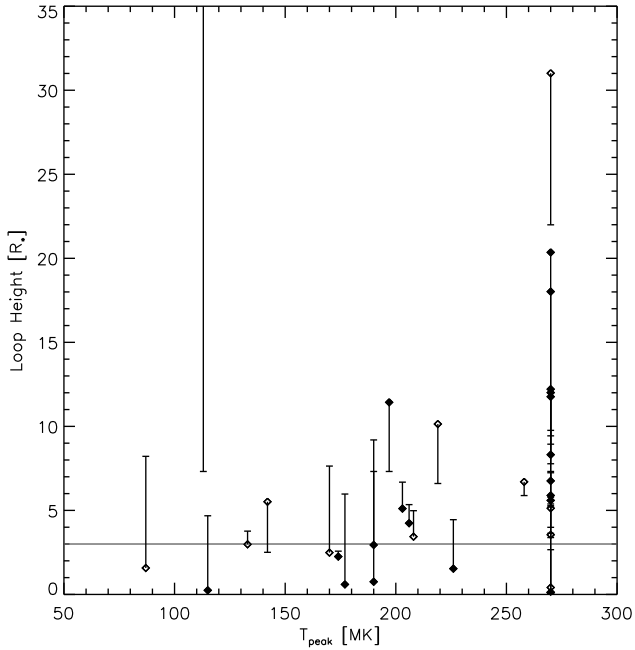}
\end{center}
\caption{A summary of the results of \citet{Favata:2005}: inferred loop height as a function of measured loop temperature.
Vertical lines connect the loop height to that star's corotation radius. We have scaled the figure for ease of viewing; 
the tallest loop (its vertical corotation line shown) is $\sim$63 R$_{*}$ in height. Filled points denote cases 
in which any parameters in calculation of the corotation radius were assumed (i.e., any values of R$_{*}$, M$_{*}$, or 
P$_{rot}$ not drawn from surveys of the ONC; fiducial values were used as discussed in section \ref{SS-fiducial}). The 
grey horizontal line at 3 R$_{*}$ indicates the location of the source 
surface (the extent of the closed coronal magnetic field) in our models. There is a pileup of points at 270 MK 
because, to be conservative, the authors adopted a maximum best-fit temperature and applied it to all cases above 
a given threshold (beyond this temperature, the fits to their X-ray data are statistically equivalent, and a higher 
temperature simply yields a shorter loop).}
\vspace{-15pt}
\label{F-f05loops}
\end{figure}

Observations of T Tauri Stars (TTS) have revealed the presence of magnetic structures much larger than those seen on 
our Sun. Indeed, analysis of these structures has invoked solar flare-based models: to infer the length scales of 
confining loops, \citet{Reale:1997} developed the uniform cooling loop (UCL) model, linking solar flare X-ray decay 
slopes to spatially resolved lengths of the loops confining the emitting gas. The UCL model was applied to data from 
the Chandra Orion Ultradeep Project \citep[COUP,][]{Getman:2005}, and it was discovered that evidently the post-flare 
loops of TTS can reach up to multiple stellar radii in extent \citep[][results summarised in Figure \ref{F-f05loops}]{Favata:2005}.
This result is surprising: T Tauri coronae are typically believed to be compact, as evidenced by the common detection of 
rotationally modulated coronal X-ray emission \citep{Flaccomio:2005}. X-ray bright regions are generally thought to be 
confined within the corona, with loops generally not reaching more than a stellar radius in height.

In the mechanical equilibrium model of \citet{Jardine:2005}, prominences form after reconnection occurs in the open field, 
and it is a straightforward matter to show that cool, dense loops can reach heights well beyond corotation (the radius at 
which the orbital velocity equals the stellar rotational velocity) and the closed stellar coronal field. Hot loops such as 
those inferred by the COUP, however, appear to need the support of the stellar field, and thus should be limited to heights 
within the closed coronal magnetic field: a fraction of the heights inferred by UCL models. 
It was initially suggested that perhaps these hot, large-scale loops are anchored to circumstellar disc material for 
stability. Testing this idea, followup studies of the COUP objects that searched for discs were unable to find conclusive 
evidence for dusty disc material within reach of the loops \citep[e.g.,][]{Getman:2008,Aarnio:2010}. Furthermore, the 
apparent anti-correlation of large-scale loops and close-in disc material suggests it is the absence of inner disc 
material that allows the loops to reach greater extents than they would were close-in disc material present 
\citep[as suggested by][close-in circumstellar discs could strip away these stars' outer coronae]{Jardine:2006}. 
Potentially, this indicates some missing physics in earlier models, and it raises 
the question of how such large loops remain stable for the duration of the X-ray flare decay phase (in some cases, multiple 
rotation periods).

The potential ramifications of highly energetic events on stellar and circumstellar evolution are far-reaching: 
large-scale magnetic structures, in their formation and eventual disruption, could regulate stellar angular momentum 
evolution as well as circumstellar disc evolution, and thus planet formation. The combination of the high frequency 
with which large-scale magnetic structures have been inferred and the great heights at which they form above the stellar 
surface indicates circumstances are ideal for the generation of substantial torques opposing stellar
rotation. Large-scale magnetic reconnection, in addition to shedding mass and angular momentum, also bombards 
circumstellar material with high energy photons and particles. This could potentially play a pivotal role in planet 
formation by stimulating grain growth \citep[cf. flash-heated chondrule formation,][]{Miura:2007}.
High energy radiation and particle fluxes can influence disc chemistry 
\citep[cf., solar nebula isotopic abundances,][]{Feigelson:2002} and structure: 
\citet{Glassgold:1997} showed that hard stellar X-rays could potentially ionise the disc sufficiently for 
accretion to occur via magnetohydrodynamic instability. The large heights inferred for these loops would put X-ray 
emitting plasma closer to the disc, reducing the column depth through which the hard X-rays would have to travel 
before ionising disc material.

Given these potential impacts of large scale magnetic loops (and their destabilization) on early star and disc evolution, 
we aim to ascertain whether the existence of hot, large, magnetic loops, as inferred from the analysis of the brightest, 
most energetic events in X-ray surveys of TTS, is physically plausible. To do this, we address an evident buoyancy 
defecit in the previous models by including a stellar wind in the model physics. We do not distinguish between the 
single loop UCL model or the possibility of multiple loop events; we only seek to demonstrate that it is possible to 
find hot loops beyond the typical ``quiescent'' X-ray emitting corona, and in some cases, beyond corotation.

\section{Mechanical Equilibrium Model}

Cool, extended prominences have been observed in abundance on the young rapid rotator AB Dor, ranging in height from 
2-5~R$_{*}$, well above the corotation radius of 1.7~R$_{*}$ \citep{Cameron:1989}, and slingshot prominences much like 
those seen on the Sun have been observed on TTS \citep[e.g.,][]{Massi:2008,Skelly:2008}. 
From a theoretical perspective, other groups \citep[][hereafter JCC91 and JvB05, respectively]{Jardine:1991,Jardine:2005}
have assessed the stability of cool loops, determining where in parameter space mechanical equilibria can be found.
Loops found beyond the closed coronal field and possibly even beyond corotation are likely embedded in the 
stellar wind (these loops could form via reconnection of open field lines in the wind; see JvB05, Figure 2).

This work builds upon the prescription of JCC91 and JvB05, save our addition of a stellar wind (Section \ref{S-wind}). 
We list below the four major forces our model accounts for (and the parameters upon which each depends) that dictate 
the height and shape of a magnetic loop:
\begin{enumerate}
\item{} Gravity (M$_{*}$, R$_{*}$, $\omega$),
\item{} Buoyancy ($\rho$, $g$, $T$, $\partial p/\partial y$),
\item{} Magnetic pressure gradient (assumed field structure),
\item{} Magnetic tension (footpoint separation, external field strength and structure).
\end{enumerate}
A magnetic loop is in equilibrium when the pressure gradients internal and external to the loop, magnetic tension, 
and gravity are evenly balanced: the loop is stable, neither expanding nor contracting. We describe the general mechanical 
equilibrium model, and the addition of a stellar wind term. In Section \ref{S-toymodel}, the case of a simple TTS is 
demonstrated.

\subsection{Equilibrium Conditions}

We incorporate a number of fairly standard, simplifying assumptions in our 2D model. First, 
the arcade in which the loop is embedded is not twisted \citep[i.e. 
no shear;][found, in the solar case, that including twist in the flux tube only serves to 
reduce the maximum loop footpoint separation]{Browning:1984}. Secondly, the external field is 
potential (i.e. current free) and closed up to a height y$_s$; beyond this point, 
the source surface, the external field is open. The source surface sets, then, 
the maximum height quiescent coronal loops can reach. The magnetic loop itself 
is assumed to be narrow, its width less than the length scale of the external field, and it is 
isothermal. Finally, the loop is treated as non-disruptive of its surroundings. The equation of 
motion for plasma in the corona is:
\begin{equation}\label{eqn-pbal}
0 = - \bf{\nabla} \left( \mathrm{p} + \frac{\mathrm{B}^2}{\mathrm{8}\pi}\right) + \frac{\mathrm{1}}{\mathrm{4}\pi}(\bf{B}\cdot\bf{\nabla})\bf{B} + \rho \bf{g},
\end{equation}
where the first term expresses the plasma and magnetic pressure gradients, the second term 
represents the magnetic tension, and the final term is the balance of gravitational and 
centrifugal forces. In all the following derivation, we refer to quantities inside the loop as 
``internal'', and coronal properties as ``external'' (variable subscripts $i$ and $e$, respectively).

\begin{figure}
\begin{center}
\includegraphics[width=85mm,clip=true,trim=30. 0 10. 20.]{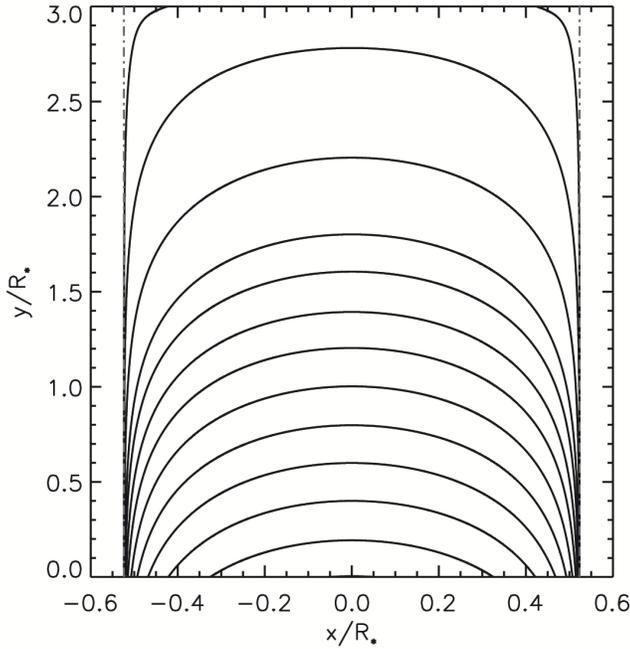}
\end{center}
\caption{A graphical representation of the form of the external field (see equation \ref{eqn-bext}). Our chosen source 
surface height, where the magnitude of the field is zero, is 3.0 R$_{*}$; beyond this height, the field is open. 
Vertical lines indicate the width of the arcade ($|x| = \pi/k$, where here we choose $k=3$ so the loop covers 
60$^\circ$ in longitude about the stellar equator).}
\label{F-bext}
\end{figure}

We choose a simple external field which is potential and two dimensional:
\begin{equation}\label{eqn-bext}
B_{ex} + \mathrm{i}B_{ey} = B_0 (e^{2\mathrm{i}kz}+e^{-2ky_s})^{1/2},
\end{equation}
where $z = x + $i$y$ (see Figure \ref{F-bext}, illustration of field) and $B_0$ is the field strength 
at the stellar surface. Rewriting the Lorentz force in terms of the current density, $\textbf{j}$, we have, 
\begin{equation}\label{eqn-aeom}
\bf{\nabla} \left( \frac{\mathrm{B}^2}{\mathrm{8}\pi} \right) + \frac{\mathrm{1}}{\mathrm{4}\pi} (\bf{B}\cdot\bf{\nabla})\bf{B} = \bf{j} \times \bf{B}.
\end{equation}
Since the Lorentz force is zero in the direction of the field (hence $\textbf{j}\times\textbf{B}=$\textbf{0}), 
we can combine equations (\ref{eqn-pbal}) and (\ref{eqn-aeom}) to obtain an expression for pressure balance along the loop:
\begin{equation}\label{eqn-EoM}
\frac{dp}{ds} = \rho g_s,
\end{equation}
where $g_s =\bf{g \cdot B/|B|}$ is the component of gravity along the loop. In a Cartesian coordinate system where $\bf{g} =$ (0,g), 
this reduces to:
\begin{equation}
\frac{dp}{dy} = \rho g(y).
\end{equation}

We assume the plasma along a given loop is in hydrostatic equilibrium, so we solve for it as: 
\begin{equation}\label{eqn-pinternal}
p = p_0 \mathrm{exp} \left( \frac{\mu m_H}{k_B T} \int_0^y g dy \right),
\end{equation}
where $p_{0}$ is the loop base pressure (i.e., where y$=$0). 

We note that pressure balance across the loop requires: 
\begin{equation}\label{eqn-pressbal}
B_i^2 = B_e^2 + 8\pi(p_e-p_i).
\end{equation}

Following the derivation and boundary conditions of JvB05, we consider gradients normal to the field by taking 
the scalar product of $\bf{\hat{n}}$ and both sides of equation (\ref{eqn-aeom}), arriving at an expression that 
governs the shape of the loop,
\begin{equation}\label{eqn-loopshape}
\left(y'\frac{\partial}{\partial x}-\frac{\partial}{\partial y}\right)\frac{B_i^2}{2} = -\frac{y''}{1+(y')^2}B_i^2.
\end{equation}

The loop shape (given by $y(x)$) is therefore determined by the requirement that the Lorentz force is zero and 
hence that the magnetic pressure gradients are balanced by the magnetic tension. At the loop summit, magnetic 
tension acts directly downwards, so we must have $\partial{B^2}/\partial{y} < 0$. The behaviour of $B_i^2$, however, 
is determined through pressure balance (equation \ref{eqn-pressbal}) by the nature of the external magnetic field and the 
variation of the internal and external plasma pressures. The requirement that $\partial{B^2_i}/\partial{y} < 0$
reduces to:
\begin{equation}\label{eqn_balance}
\frac{1}{8\pi} \frac{\partial B_e^2}{\partial y} < (\rho_i - \rho_e)g.
\end{equation}

From equation (\ref{eqn_balance}), we begin to see some necessary physical conditions for stable loops. The maximum 
loop height is constrained by how buoyancy (i.e., the gradient of the pressure difference, $\partial(p_{e}-p_{i}$)/$\partial y$) 
depends on height. In the absence of a wind, the gas pressure external to the loop falls with increasing height until a 
turning point is reached at corotation: at this radius, centrifugal forces dominate over gravitational, and the buoyancy 
changes sign. If equation (\ref{eqn_balance}) is satisfied, the loop can be supported by the external field; if not 
satisfied, the loop will lose equilibrium.

Meeting these conditions becomes a problem in the T Tauri case, as densities measured in hot loops are greater than ambient 
coronal densities, the internal temperatures are likely greater than the external temperatures, yet the inferred heights put 
them beyond corotation.  By our previous understanding of the physics, this combination of parameters would not result in 
a mechanically stable loop within a hydrostatic corona.

For cool prominences, equilibria existed beyond corotation only for negatively buoyant 
loops; 
in addition to the loop density being greater than its surroundings, the plasma $\beta$ (ratio of gas
to magnetic pressure) is sufficiently large, but less than unity. We can adjust the plasma $\beta$ in 
the previous models and find solutions for tall, hot loops, but the value of $\beta$ required for solutions 
beyond corotation is unphysical; for $\beta$ values consistent with measured parameters on TTS, hot 
loops are positively buoyant to large distances. To provide additional upward force to support the 
hot loop, we add a stellar wind external to the loop. In this case, hot loops can reach a greater 
range of heights because the wind term extends the range over which buoyancy is negative. 

\subsection{Including a Stellar Wind}\label{S-wind}

In order to maintain stability of very extended, hot loops, the external plasma pressure must decrease 
approaching and passing corotation in order to allow the loop to maintain negative buoyancy at 
large heights. Far enough from the star, the external magnetic pressure becomes negligible, and 
so the internal field strength (and thus the loop shape) is determined by the plasma pressure 
difference. Now, we include an additional term to include a lowering of the pressure due to 
the flow of a stellar wind along the open external field in which the magnetic loop is embedded; 
this will provide more upwards force on the loop, potentially allowing greater heights to be obtained 
by loops with a wider range of parameters (consistent with inferred parameters of TTS magnetic loops).

We have added to the basic formulation of JvB05 a simple, Parker wind \citep{Parker:1964a}. Assuming 
an inviscid parcel of gas is corotating with a star, the total force on that parcel is:
\begin{equation}
\mathbf{F} = - \nabla p + \rho \mathbf{g} + \mathbf{j} \times \mathbf{B}.
\end{equation}
As this is an expression of Newton's second law, it can be rewritten as follows:
\begin{equation}
\mathbf{F} = \rho \frac{D \mathbf{v}}{D t},
\end{equation}
where $\frac{D}{D t}$ is the substantive derivative,
\begin{equation}
\frac{D}{D t} = \frac{\partial}{\partial t} + (\mathbf{v} \cdot \nabla).
\end{equation}
We will assume here both a potential field and a steady flow, so the momentum equation along the 
direction of the loop (i.e., parallel to the direction of the field: 
$\mathbf{\hat{s}}=\mathbf{B}/|\mathbf{B}|$) is:
\begin{equation}
\rho \mathbf{\hat{s}} \cdot (\mathbf{v} \cdot \nabla) \mathbf{v} 
   = - \mathbf{\hat{s}} \cdot \nabla \mathbf{p} + \rho \mathbf{g} \cdot \mathbf{\hat{s}}.
\end{equation}

Under the assumption of an isothermal gas, density and pressure are related by the local 
sound speed of the plasma, $\rho = \frac{p}{c_s^2}$. Conservation of mass requires that the flow 
across a given cross sectional area remain constant (i.e., $\rho v A = \mathrm{constant}$,) and 
similarly the conservation of magnetic flux dictates no sinks or sources of magnetic field 
($B A = \mathrm{constant}$).

Bearing these relationships in mind, the momentum equation along the loop can be re-written as:
\begin{equation}\label{eqn_mach}
\frac{\partial}{\partial s} \left( \frac{v^2}{2} \right) = 
   \frac{-c_s^2}{p} \frac{\partial p}{\partial s} + \mathbf{g} \cdot \mathbf{\hat{s}}.
\end{equation} 

Integrating equation (\ref{eqn_mach}) from the loop base upward, we obtain the following expression 
relating the external field and pressure, and $M = \frac{v_0}{c_s}$ the initial sonic Mach number 
of the flow:
\begin{equation}\label{eqn-pb}
0 = \mathrm{ln}(p_e) + \frac{1}{2} M^2 \left( \frac{B_e}{p_e} \right)^2 - \frac{1}{2} M^2 
              - \frac{1}{c_s^2} \int g dy.
\end{equation}
The external pressure and field, $p_e$ and $B_e$, are scaled by their base values 
($p_0$ and $B_0$, respectively).

If the initial Mach number is 0, the momentum equation reduces to the hydrostatic case (equation \ref{eqn-pinternal}).
The only physical solution of equation (\ref{eqn-pb}) is the transonic solution, which gives a 
slow, sub-sonic speed close to the star, passes through a sonic point (where the flow velocity equals 
the local sound speed), and is supersonic at greater distances from the star \citep[see Figure 1 of ][]{Parker:1965}.
To illustrate how including a stellar wind can impact the maximum possible loop height, we consider 
the specific case of a TTS below.

\section{Specific Case: a T Tauri Star}\label{S-toymodel}

We have applied the mechanical equilibrium model plus wind to a generic, simple T Tauri star. The basic 
outline of the model and analyses over robust parameter spaces are described fully in previous work 
\citep[JCC91;][JvB05]{Unruh:1997}; here we give full attention to the new addition of a stellar wind term 
and its effects on the physics of the model. First, we discuss possible equilibria for hot loops embedded 
within the closed corona; second, we assess maximum heights for loops extending beyond the source surface. 

\subsection{Fiducial T Tauri Parameters}\label{SS-fiducial}

As input parameters to the model, we require stellar mass, radius, and rotation period; we have adopted 
representative mean values from the Orion Nebula Cluster from the study of \citet{LAH:1997}. The mean mass 
of members (with 50\% membership probabilities or greater) is 0.55 M$_{\odot}$, while the mean radius is 
1.7 R$_{\odot}$. Our fiducial TTS rotation period of 5.3 days is derived from the assembled rotation period 
data of \citet{Getman:2005}. We adopt a wind temperature of 2 MK, consistent with T Tauri wind observations 
\citep{Dupree:2005}. Internal loop temperatures have been taken from the range of values measured for the 
COUP ``superflares'' inferred by \citet{Favata:2005}; these vary from $\sim$50-200 MK (see Figure \ref{F-f05loops}).
We set the initial Mach number to 0.05. 
For the extent of the closed coronal field, we show our results for y$_s =$ 3~R$_{*}$, which is within 
the corotation radius (5.2~R$_{*}$). It is unclear for TTS how far from the stellar surface the closed 
corona may extend, but as we note in Section \ref{S-discus}, our results do not depend strongly on the 
choice of source surface height (see Section \ref{S-shapes} for further discussion of y$_{s}$).

\subsection{Pressure and Buoyancy}

For the hydrostatic case, the functional form of the external plasma pressure is exponentially decreasing 
close to the star, with the argument of the exponential function dependent upon gravitational and centrifugal 
forces. In the case of a T Tauri star specifically, given the same mass and rotation period as a main-sequence 
counterpart but twice the radius, the effective gravity is approximately four times weaker, thereby 
extending the distances to physically important heights. The corotation radius for the TTS is a factor 
of two closer, and the difference in pressure (i.e., $p_e-p_i$) remains below 
the base pressure difference to a distance approximately three times closer than the main sequence star. 
The corotation radius is important because it is where pressure difference and the buoyancy change sign.

\begin{figure}
\includegraphics[width=85mm,clip=true,trim=30. 0 10. 20.]{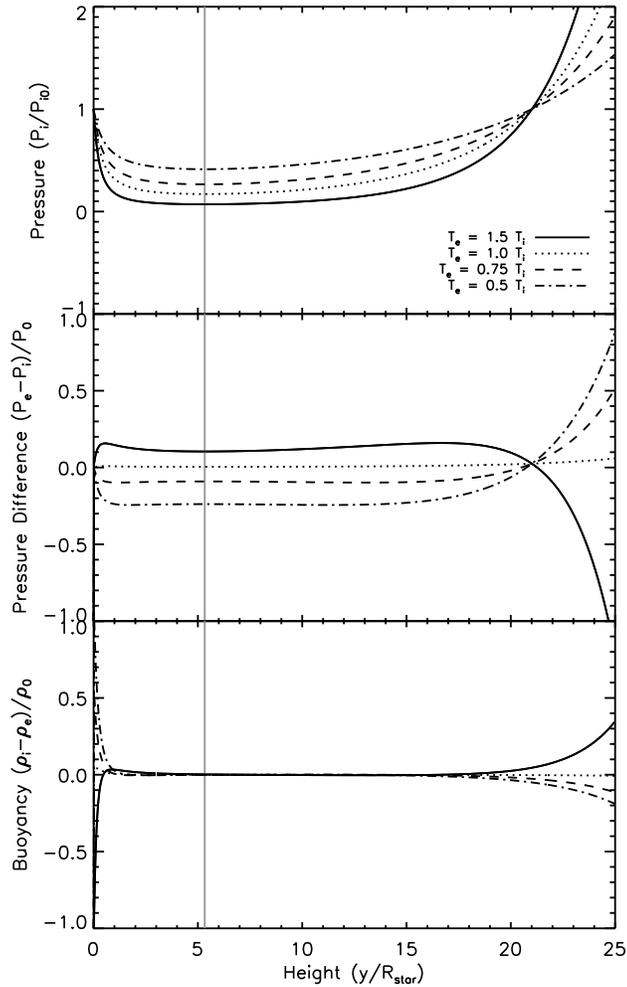}
\caption{
Pressure, pressure difference, and buoyancy as a function of height (upper, middle, and lower panels, 
respectively). The external plasma temperature is 2 MK, and temperature ratios (T$_{e}$/T$_{i}$) shown are 
1.5 (solid line), 1.0 (dotted line), 0.75 (dashed line), and 0.5 (dot-dashed line). The corotation radius is 
located at 5.2 R$_{*}$ (grey vertical lines); at 20.2 R$_{*}$ the internal pressure equals its base value, 
and the pressure difference changes sign.}
\label{F-press}
\end{figure}

\begin{figure}
\includegraphics[width=85mm,clip=true,trim=60.0 30.0 20. 20.]{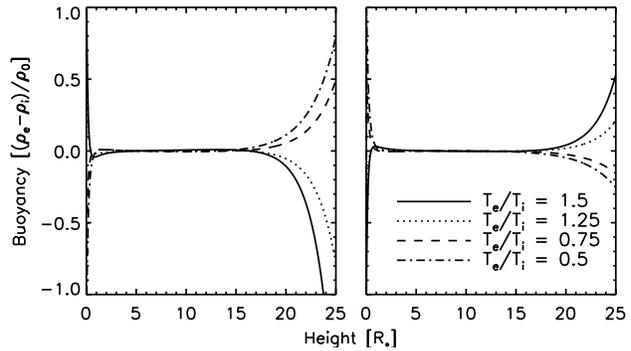}
\caption{A comparison of loop buoyancy using the physics of JvB05 (left panel) and our present model including a stellar 
wind (right panel). For our fiducial T Tauri parameters, under the previous model physics, loops much hotter than 
their surroundings were negatively buoyant only fairly close to the stellar surface, within the source surface.}
\label{F-comparebuoy}
\end{figure}

We choose a base pressure and temperature ratio, and with these we solve equation (\ref{eqn-pb}) for the 
equilibrium shape of the loop. 
In Figure \ref{F-press} we show, for the parameters outlined above, the internal pressure, pressure 
difference ($p_{e}-p_{i}$), and buoyancy ($\partial(p_{e}-p_{i})$/$\partial y$) as a function of height 
for various temperature ratios $T_{e}/T_{i}$. It is important to note the heights at which the loop's buoyancy is 
negative when including the stellar wind; in Figure \ref{F-comparebuoy} we illustrate how our addition 
of the stellar wind affects loop buoyancy. Where cool prominences had negative buoyancy out to great 
heights previously (Figure \ref{F-comparebuoy}, left panel, solid/dotted lines), we now see that 
including the wind means hot loops have negative buoyancy at large distances from the stellar surface 
(Figure \ref{F-comparebuoy}, right panel, dashed/dot-dashed lines).

\subsection{Loop Shapes, Maximum Heights}\label{S-shapes}

The critical parameter we are altering in this analysis is the external pressure (and thus the pressure 
balance of the loop with its environs).
We have improved on previous models that assumed a hydrostatic external plasma pressure to include the effect 
of a wind. The height of the source surface (the height beyond which all field lines are open) is not a 
well-constrained parameter, 
as interpretations of multi-wavelength observations imply conflicting values. Solar observations show a 
white light corona which extends to greater heights than the X-ray bright corona. In the stellar case, densities 
derived from X-ray emission measures suggest more compact coronae, while the lack of apparent rotational 
modulation in some cases, as well as the formation of large-scale prominences, indicates the X-ray coronae 
are extended. Our choice reflects a field more extended than main-sequence, solar type stars, but less so 
than a purely dipolar field \citep[based on analysis of the relationship between field complexity and the 
observed coronal X-ray emission measure, ][]{Jardine:2006}.

\begin{figure}
\includegraphics[width=85mm,clip=true,trim=30. 0 10. 20.]{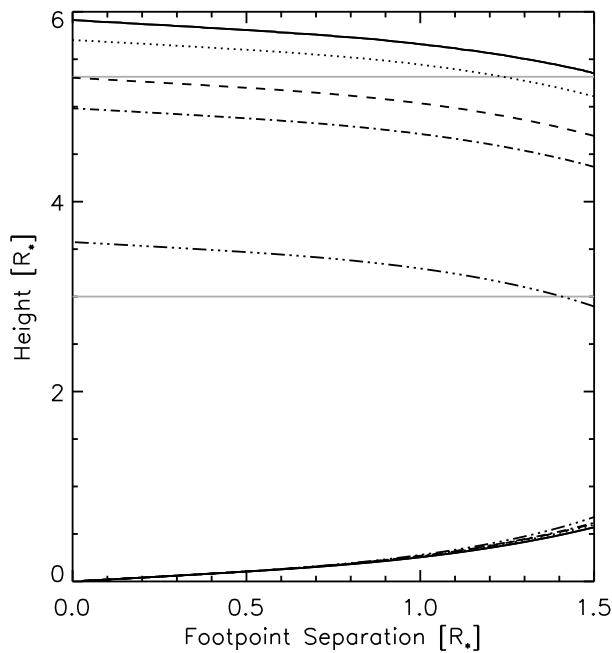}
\caption{
Loop heights as a function of footpoint separation. Our loop temperatures probe the parameter space illustrated in 
Figure \ref{F-f05loops}: for loop temperatures from 30-200 MK, we find loops with heights above the closed coronal 
field \citep[consistent with the inferences of ][]{Favata:2005}. The loop temperatures are: 30 MK (solid line), 
40 MK (dotted line), 50 MK (dashed line), 60 MK (dot-dashed line), and 200 MK (triple dot-dashed line). Grey, horizontal 
lines denote the source surface (extent of closed coronal field) and 
corotation radii at 3 R$_{*}$ and 5.2 R$_{*}$, respectively. T$_{e}$ is 2 MK, the plasma 
$\beta$ is 0.05, and the stellar parameters used are as described in Section \ref{S-toymodel}. For all loops shown 
here, p$_{i}$ is $\sim$10-20\% lower than p$_{e}$. Footpoint separations beyond 
1.05 R$_{*}$, though mathematically possible, are non-physical solutions, as we have defined the arcade width to 
be $x = \pi/k$, where $k=3$. For a fairly narrow 
range of loop temperatures, we find a wide range of possible maximum loop heights, the coolest loops reaching 
just beyond corotation, and the hottest loops confined within the closed corona.}
\label{F-shape_ys3}
\end{figure}

In Figure \ref{F-shape_ys3}, we show the shape of hot loops when the source surface is placed 3 R$_{*}$ 
from the stellar surface. Within the closed corona, the pressure gradient of the external field is 
directed upward, providing support for the loop against the downward force of magnetic tension. This set 
of solutions is effectively identical to those found for hot loops by JvB05; we too find that equilibrium 
solutions can be found for both negatively and positively buoyant loops out to the source surface.
The coolest of the loops plotted (30 MK and 40 MK) are able to reach heights above the source surface and 
above corotation due to the extra support of the wind. 

\section{Conclusions}\label{S-discus}

We have addressed the issue of mechanical equilibrium for large, hot loops. Since cases have been found in 
which hot loops several stellar radii in height seem to not be disc-supported, we propose that these loops 
reach mechanical equilibria beyond the closed coronal magnetic field due to the stellar wind. We have shown 
that, for a generic T Tauri star, our equilibrium loop heights are consistent with the inferred loop lengths 
for the most powerful flares observed by the COUP. We also recover solutions for hot loops within the range 
of heights below the source surface anticipated by previous models, suggesting that large and small stable 
post-flare loops can occur under similar conditions. In the physics of previous models, which 
were hydrostatic and did not consider the role of stellar winds, loops hotter than the 
quiescent corona were confined to heights within the closed coronal field, needing its support. Our present 
work indicates that, given a stellar wind, hot loops can be found at heights well beyond the source surface, 
and even beyond corotation.

Young stars are known to be very magnetically active in the T Tauri phase, with frequent flaring and high 
X-ray luminosities. The apparent abundance of large scale loops, favourable conditions for their formation 
in TTS coronae and winds, and potential for consequence on pre-main sequence stellar (and circumstellar) 
evolution makes understanding these large-scale loops very important. Having verified that such large-scale 
magnetic structures can indeed exist, interesting subsequent work will be to determine how the loops' 
disruption can impact circumstellar disc material as well as the star itself. Large-scale disruptions of 
this nature could effect stellar rotational evolution and activity as well as circumstellar disc evolution 
and planet formation.

\section*{Acknowledgments}

ANA acknowledges support for this work from NSF AST-0808072 (P.I. K. Stassun) and a STFC visitor grant. SGG 
is supported by NASA grant HST-GO-11616.07-A. JL acknowledges the support of an STFC studentship.

\bibliographystyle{mn2e}
\bibliography{ttloops}

\begin{thebibliography}{}

\bibitem[\protect\citeauthoryear{{Aarnio}, {Stassun} \& {Matt}}{{Aarnio}
  et~al.}{2010}]{Aarnio:2010}
{Aarnio} A.~N.,  {Stassun} K.~G.,    {Matt} S.~P.,  2010, \apj, 717, 93

\bibitem[\protect\citeauthoryear{{Browning} \& {Priest}}{{Browning} \&
  {Priest}}{1984}]{Browning:1984}
{Browning} P.~K.,  {Priest} E.~R.,  1984, \solphys, 92, 173

\bibitem[\protect\citeauthoryear{{Collier Cameron} \& {Robinson}}{{Collier
  Cameron} \& {Robinson}}{1989}]{Cameron:1989}
{Collier Cameron} A.,  {Robinson} R.~D.,  1989, \mnras, 236, 57

\bibitem[\protect\citeauthoryear{{Dupree}, {Brickhouse}, {Smith} \&
  {Strader}}{{Dupree} et~al.}{2005}]{Dupree:2005}
{Dupree} A.~K.,  {Brickhouse} N.~S.,  {Smith} G.~H.,    {Strader} J.,  2005,
  \apjl, 625, L131

\bibitem[\protect\citeauthoryear{{Favata}, {Flaccomio}, {Reale}, {Micela},
  {Sciortino}, {Shang}, {Stassun} \& {Feigelson}}{{Favata}
  et~al.}{2005}]{Favata:2005}
{Favata} F.,  {Flaccomio} E.,  {Reale} F.,  {Micela} G.,  {Sciortino} S.,
  {Shang} H.,  {Stassun} K.~G.,    {Feigelson} E.~D.,  2005, \apjs, 160, 469

\bibitem[\protect\citeauthoryear{{Feigelson}, {Garmire} \&
  {Pravdo}}{{Feigelson} et~al.}{2002}]{Feigelson:2002}
{Feigelson} E.~D.,  {Garmire} G.~P.,    {Pravdo} S.~H.,  2002, \apj, 572, 335

\bibitem[\protect\citeauthoryear{{Flaccomio}, {Micela}, {Sciortino},
  {Feigelson}, {Herbst}, {Favata}, {Harnden} Jr. \& {Vrtilek}}{{Flaccomio}
  et~al.}{2005}]{Flaccomio:2005}
{Flaccomio} E.,  {Micela} G.,  {Sciortino} S.,  {Feigelson} E.~D.,  {Herbst}
  W.,  {Favata} F.,  {Harnden} Jr. F.~R.,    {Vrtilek} S.~D.,  2005, \apjs,
  160, 450

\bibitem[\protect\citeauthoryear{{Getman}, {Feigelson}, {Grosso},
  {McCaughrean}, {Micela}, {Broos}, {Garmire} \& {Townsley}}{{Getman}
  et~al.}{2005}]{Getman:2005}
{Getman} K.~V.,  {Feigelson} E.~D.,  {Grosso} N.,  {McCaughrean} M.~J.,
  {Micela} G.,  {Broos} P.,  {Garmire} G.,    {Townsley} L.,  2005, \apjs, 160,
  353

\bibitem[\protect\citeauthoryear{{Getman}, {Feigelson}, {Micela}, {Jardine},
  {Gregory} \& {Garmire}}{{Getman} et~al.}{2008}]{Getman:2008}
{Getman} K.~V.,  {Feigelson} E.~D.,  {Micela} G.,  {Jardine} M.~M.,  {Gregory}
  S.~G.,    {Garmire} G.~P.,  2008, \apj, 688, 437

\bibitem[\protect\citeauthoryear{{Glassgold}, {Najita} \& {Igea}}{{Glassgold}
  et~al.}{1997}]{Glassgold:1997}
{Glassgold} A.~E.,  {Najita} J.,    {Igea} J.,  1997, \apj, 480, 344

\bibitem[\protect\citeauthoryear{{Hillenbrand}}{{Hillenbrand}}{1997}]{LAH:1997}
{Hillenbrand} L.~A.,  1997, \aj, 113, 1733

\bibitem[\protect\citeauthoryear{{Jardine} \& {Collier Cameron}}{{Jardine} \&
  {Collier Cameron}}{1991}]{Jardine:1991}
{Jardine} M.,  {Collier Cameron} A.,  1991, \solphys, 131, 269

\bibitem[\protect\citeauthoryear{{Jardine}, {Collier Cameron}, {Donati},
  {Gregory} \& {Wood}}{{Jardine} et~al.}{2006}]{Jardine:2006}
{Jardine} M.,  {Collier Cameron} A.,  {Donati} J.-F.,  {Gregory} S.~G.,
  {Wood} K.,  2006, \mnras, 367, 917

\bibitem[\protect\citeauthoryear{{Jardine} \& {van Ballegooijen}}{{Jardine} \&
  {van Ballegooijen}}{2005}]{Jardine:2005}
{Jardine} M.,  {van Ballegooijen} A.~A.,  2005, \mnras, 361, 1173

\bibitem[\protect\citeauthoryear{{Massi}, {Ros}, {Menten}, {Kaufman
  Bernad{\'o}}, {Torricelli-Ciamponi}, {Neidh{\"o}fer}, {Boden}, {Boboltz},
  {Sargent} \& {Torres}}{{Massi} et~al.}{2008}]{Massi:2008}
{Massi} M.,  {Ros} E.,  {Menten} K.~M.,  {Kaufman Bernad{\'o}} M.,
  {Torricelli-Ciamponi} G.,  {Neidh{\"o}fer} J.,  {Boden} A.,  {Boboltz} D.,
  {Sargent} A.,    {Torres} G.,  2008, \aap, 480, 489

\bibitem[\protect\citeauthoryear{{Miura} \& {Nakamoto}}{{Miura} \&
  {Nakamoto}}{2007}]{Miura:2007}
{Miura} H.,  {Nakamoto} T.,  2007, Icarus, 188, 246

\bibitem[\protect\citeauthoryear{{Parker}}{{Parker}}{1964}]{Parker:1964a}
{Parker} E.~N.,  1964, \apj, 139, 72

\bibitem[\protect\citeauthoryear{{Parker}}{{Parker}}{1965}]{Parker:1965}
{Parker} E.~N.,  1965, \apj, 141, 1463

\bibitem[\protect\citeauthoryear{{Reale}, {Betta}, {Peres}, {Serio} \&
  {McTiernan}}{{Reale} et~al.}{1997}]{Reale:1997}
{Reale} F.,  {Betta} R.,  {Peres} G.,  {Serio} S.,    {McTiernan} J.,  1997,
  \aap, 325, 782

\bibitem[\protect\citeauthoryear{{Skelly}, {Unruh}, {Cameron}, {Barnes},
  {Donati}, {Lawson} \& {Carter}}{{Skelly} et~al.}{2008}]{Skelly:2008}
{Skelly} M.~B.,  {Unruh} Y.~C.,  {Cameron} A.~C.,  {Barnes} J.~R.,  {Donati}
  J.,  {Lawson} W.~A.,    {Carter} B.~D.,  2008, \mnras, 385, 708

\bibitem[\protect\citeauthoryear{{Unruh} \& {Jardine}}{{Unruh} \&
  {Jardine}}{1997}]{Unruh:1997}
{Unruh} Y.~C.,  {Jardine} M.,  1997, \aap, 321, 177

\end{thebibliography}

\label{lastpage}

\end{document}